\documentclass[aps,prb,twocolumn,showpacs,floatfix,superscriptaddress,nofootinbib]{revtex4}
\usepackage{graphicx}
\usepackage{hyperref}
\usepackage{amssymb}
\usepackage{amsmath}
\usepackage{lmodern}
\usepackage{color}
\usepackage{empheq}
\usepackage{epstopdf}
\usepackage{amsmath}

\begin{document}

\title{The exact travelling wave solutions of a KPP equation}

\author{Eugene Kogan}
\email{Eugene.Kogan@biu.ac.il}
\affiliation{Department of Physics, Bar-Ilan University, Ramat-Gan 52900, Israel}

\begin{abstract}
We  obtain the exact analytical traveling wave solutions 
of the  Kolmogorov-Petrovskii-Piskunov equation with the reaction term belonging to the class of functions, which includes that of the (generalized) Fisher equation,
for the particular values of the waves speed. Additionally we obtain the exact analytical traveling wave solutions of the generalized Burgers–Huxley equation.

\end{abstract}

\date{\today}

\maketitle

\section{Introduction}

Fisher \cite{fisher} first introduced a nonlinear evolution equation to investigate the wave
propagation of an advantageous gene in a population. His equation also describes the
logistic growth–diffusion process and has the form
\begin{eqnarray}
\label{fi}
u_t-\nu u_{xx}=ku\left(1-\frac{u}{\kappa}\right)
\end{eqnarray}
where $\kappa(> 0)$
is the carrying capacity of the environment. The term 
$f(u)=ku(1-u/\kappa)$ represents
a nonlinear growth rate which is proportional to $u$ for small $u$, but decreases as $u$
increases, and vanishes when $u = \kappa$. 
Equation (\ref{fi}) arises in many physical,
biological, and chemical problems involving diffusion and nonlinear growth.
It is convenient to introduce the nondimensional
quantities 
\begin{subequations}
\begin{alignat}{4}
x^*&=x\sqrt{k/\nu},\\
t^*&=kt,\\
u^*&=u/\kappa,
\end{alignat}
\end{subequations}
so after
dropping the asterisks, equation (\ref{fi}) takes the
nondimensional form
\begin{eqnarray}
\label{85}
u_t-u_{xx}=u\left(1-u\right).
\end{eqnarray}

In the same year (1937) as Fisher, Kolmogorov, Petrovsky and Piskunov  (KPP) \cite{kolmogorov} introduced the more general reaction-diffusion equation, which in the nondimensional form can be presented as
\begin{eqnarray}
\label{kolm}
u_t-u_{xx}=F(u),
\end{eqnarray}
where 
$F$ is a sufficiently smooth function with the properties that 
$F(0)=F(1)=0$, $F'(0)>0$ and $F'(1)<0$.

Fisher-KPP equation belongs to the class of reaction-diffusion equations: in fact, it is one of the simplest semilinear reaction-diffusion equations,  which can exhibit traveling wave solutions that switch between equilibrium states given by 
$F(u)=0$. Such equations occur, e.g., in ecology, where it describes, like it was mentioned above, the population growth, physiology, combustion, crystallization, plasma physics, and in general phase transition problems.

Equation (\ref{kolm}) admits traveling
wave solutions
\begin{eqnarray}
u(x,t)=u(\xi),\hskip .5cm \xi=x-ct,
\end{eqnarray}
where  $c$ is the wave speed (measured in the units of $\sqrt{\nu/k}$), and $u(\xi)$ satisfies the
equation
\begin{eqnarray}
\label{86}
u_{\xi\xi}+c u_{\xi}=-F(u)
\end{eqnarray}
with the boundary conditions
\begin{subequations}
\label{bc}
\begin{alignat}{4}
\lim_{\xi\to-\infty}u(\xi)&=1,\label{coco}\\
\lim_{\xi\to+\infty}u(\xi)&=0.\label{cococo}
\end{alignat}
\end{subequations}

Equation (\ref{86}) has a simple mechanical interpretation. It describes motion of the Newtonian particle with the mass 1 ($u$ being the coordinate of the particle and $\xi$ - the time) in the potential well $U(u)$ defined by the equation 
\begin{eqnarray}
\frac{dU}{du}=F(u).
\end{eqnarray}
The motion is with friction, and the coefficient of friction is equal to $c$.
 The point $u=0$ is the point of stable equilibrium, and the point $u=1$ - unstable. 

Equation (\ref{86}) is a nonlinear nonintegrable differential equation. 
The Cauchy problem cannot be solved
by the inverse scattering transform for this equation and the problem of finding some exact
solutions for this equation is an important task.
Several methods of obtaining such solutions has been developed. We will mention here a few of them:
 the tanh-function method \cite{4,5,6}, the Jacobi elliptic function expansion
method \cite{8}, the simplest equation method \cite{11,12}, the Exp-function method \cite{14}, the 
$G'/G$ expansion
method \cite{15,16,17,18}, the Kudryashov method \cite{19}, and the attached flow method
\cite{iacobescu,polyanin}.

In this paper we will use the method of exclusion of the independent variable \cite{benguria,rosu,kogan3,kogan4,kogan5}, which allows to reduce the problem of integration of the second order ODE to the problem of integration of the first order ODE,  to obtain the exact solutions of Eq. (\ref{86}). Thus we were able to obtain the exact solutions of the KPP equation which to the best of our knowledge were not known before.
Additionally we obtain the exact analytical traveling wave solutions of the generalized Burgers–Huxley equation. 

\section{The ODE which doesn't contain explicitly the independent variable}
\label{two}

Equation (\ref{86})   doesn't contain explicitly the independent variable $\xi$. This prompts the idea   to
consider $u$ as the new independent variable and
\begin{eqnarray}
\label{cot}
p=u_{\xi}
\end{eqnarray}
as the new dependent variable.
In the new variables
(\ref{86})  takes the form  of Abel equation of the second kind   \cite{polyanin}.
\begin{eqnarray}
\label{pp}
pp_{u}+cp=-F(u).
\end{eqnarray}
The boundary conditions for  (\ref{86}) were (\ref{bc}).
The boundary conditions for (\ref{pp}) are (however, see the next Section)
\begin{subequations}
\label{bmp}
\begin{alignat}{4}
p(1)&=0,\label{coco2}\\
p(0)&=0.\label{coco3}
\end{alignat}
\end{subequations}

In search of elementary solutions of (\ref{pp}), in accordance with the well-known  in mathematics principle, stating that
the more general the problem is, the easier it is to solve it, let us  generalize it to
\begin{eqnarray}
\label{hu}
pp_{u}+\gamma(u)p=-F(u),
\end{eqnarray}
where $\gamma(u)$ is some function.
Being inspired by the method of factorization \cite{gonzales},
we present the r.h.s. of (\ref{hu}) as 
\begin{eqnarray}
-F(u)=\Phi(u)\Psi(u).
\end{eqnarray}
After that we realize that there exists the particular solution of (\ref{hu})
\begin{eqnarray}
p(u)=P\Phi(u),
\end{eqnarray}
provided the unknown multiplier $P$ and the function $\gamma(u)$ satisfy equation
\begin{eqnarray}
\label{phi}
P^2\Phi_u+P\gamma(p)=\Psi(u).
\end{eqnarray}
This includes the case  of nonlinear $F(u)$ considered in Ref. \cite{an}
\begin{eqnarray}
F(U)=\left[u-A(u)\right]\left[\chi A'(u)+a\right],
\end{eqnarray}
where $A(u)$ is some function, $\chi,a$ are some numerical parameters, and $\gamma(p)=c=$ const.
In fact, taking 
\begin{eqnarray}
\Phi(u)=\left[A(u)-u\right]
\end{eqnarray}
we obtain (\ref{phi}) in the form
\begin{eqnarray}
\label{phi9}
P^2\left[A'(u)-1\right]+Pc=\chi A'(u)+a.
\end{eqnarray}
Choosing $P^2=\chi$ and $c=P+a/P$ we turn (\ref{phi9}) into identity.

Now let us be more specific and consider 
\begin{eqnarray}
\label{fff}
F(u)=u\left(1-u^n\right)\left(u^n+a\right),
\end{eqnarray}
where $n>0$ and $a\geq 0$ are some constants,
so that (\ref{86}) will take the form
\begin{eqnarray}
\label{866}
u_{\xi\xi}+c u_{\xi}=u\left(u^n-1\right)\left(u^n+a\right).
\end{eqnarray}
 Note that for $n=1/2$ and $a=1$ we recover Fisher equation. Equation (\ref{866}) with $a=1$ but $n\neq 1/2$ we may call the generalized Fisher equation \cite{kaliappan,murray,debnath}.
Starting from (\ref{866}), we obtain Eq. (\ref{pp}) in the form
\begin{eqnarray}
\label{pp2}
pp_{u}+cp=u\left(u^n-1\right)\left(u^n+a\right).
\end{eqnarray} 
We can take
\begin{eqnarray}
\Phi(u)=u\left(u^n-1\right),\hskip 1cm \Psi(u)=\left(u^n+a\right),
\end{eqnarray}
after which (\ref{phi}) takes the form
\begin{eqnarray}
\label{phio}
P^2\left[(n+1)u^n-1\right]+Pc=u^n+a.
\end{eqnarray}
We immediately realize that 
the solution of (\ref{pp2}) satisfying the boundary conditions (\ref{bmp}) 
\begin{eqnarray}
\label{pmx}
p(u)=p_0(u)\equiv Pu\left(u^n-1\right),
\end{eqnarray}
where 
\begin{eqnarray}
\label{k}
P=\frac{1}{\sqrt{n+1}},
\end{eqnarray}
exists if the speed of the wave is
\begin{eqnarray}
\label{kkk}
c=c_0\equiv P+\frac{a}{P}.
\end{eqnarray}
Substituting thus obtained $p(u)$ into (\ref{cot})  and integrating  we obtain the solution of (\ref{866})
\begin{eqnarray}
\label{mm}
u(\xi)=\frac{1}
{\left[\exp(nP\xi)+1\right]^{1/n}}.
\end{eqnarray}
For the Fisher equation from (\ref{mm}) follows the well known result \cite{ablo}
\begin{eqnarray}
\label{mmm}
u(\xi)=\frac{1}
{\left[\exp\left(\xi/\sqrt{6}\right)+1\right]^2},
\end{eqnarray}
and the condition (\ref{kkk}) takes the form
$c=5/\sqrt{6}$.

To additionally illustrate the method of integration
used in this Section, in the Appendix \ref{kudr} we obtain in
the framework of the method the exact solution 
of the generalized Duffing-Van der Pol equation.

In Appendix \ref{bh} we'll see how the results (\ref{pmx}) - (\ref{mm}) can be generalized in the case when advection is taken into account in addition to diffusion.

The connection between the existence of the elementary solution of (\ref{pp}) and the possible symmetry of the equation will be analysed in our next publication.

\section{The correct boundary conditions for Eq. (\ref{pp})}
\label{corr}

In the previous Section  
we postulated  two  boundary conditions (\ref{bmp}) for  (\ref{pp}). However,  the equation is of the first order.
The first impression is that the problem is overdetermined, and the existence of the solution of (\ref{pp}) demands
the finetuning of the l.h.s. of the equation. Being more specific,  one may think that
the  solution of (\ref{pp2})  will not exist for a general value of $c$.
However, the last statement immediately comes into contradiction with the results of the Appendix \ref{per}, where the solution of (\ref{pp2}) for arbitrary $c\gg 1$ is presented. It looks like we have a problem.

To solve this problem we have to return to Eq. (\ref{86}) with the boundary conditions (\ref{bc}). The boundary conditions  are exceptional \cite{goursat} -- $u_{\xi\xi}$  is not defined  at $\xi=\pm\infty$.
However, for  $\xi\to+\infty$ the equation asymptotically becomes
\begin{eqnarray}
\label{vv}
u_{\xi\xi}+cu_{\xi\xi}+F'(0)u=0,
\end{eqnarray}
and for $\xi\to-\infty$ -
\begin{eqnarray}
\label{vvv}
v_{\xi\xi}+cv_{\xi}-|F'(1)|v=0,
\end{eqnarray}
where $v=1-u$.

The solution of  (\ref{vvv}) is
\begin{eqnarray}
\label{can}
v=Ae^{s_1\xi}+Be^{s_2\xi},
\end{eqnarray}
where $A$ and $B$ are arbitrary constants, and $s_{1,2}$ are the roots of the characteristic equation
\begin{eqnarray}
s^2-cs-|F'(1)|=0,
\end{eqnarray}
that is
\begin{subequations}
\label{12}
\begin{alignat}{4}
s_{1}&=\frac{c+\sqrt{c^2+4|F'(1)|}}{2},\\
s_{2}&=\frac{c-\sqrt{c^2+4|F'(1)|}}{2}.
\end{alignat}
\end{subequations}
Because  of the boundary condition (\ref{coco}), $B=0$.
So this boundary condition should be specified to: For $\xi\to-\infty$
\begin{eqnarray}
\label{cand}
1-u\sim e^{s_1\xi}.
\end{eqnarray}

Now let us come to (\ref{vv}). The  characteristic equation is
\begin{eqnarray}
\label{ss}
s^2+cs+F'(0)=0,
\end{eqnarray}
that is
\begin{eqnarray}
\label{34}
s_{3,4}=\frac{-c\pm\sqrt{c^2-4F'(0)}}{2}.
\end{eqnarray}
We have either  the real roots (both being negative), the double (negative) root,
or the complex roots  (both with the negative real part). In the first case 
solution of (\ref{vv}) is
\begin{eqnarray}
\label{can2}
u=Ae^{s_3\xi}+Be^{s_4\xi},
\end{eqnarray}
for the case of double root the solution is
\begin{eqnarray}
\label{n2}
u=e^{-c\xi/2}(A+B\xi),
\end{eqnarray}
and in the complex case the solution is
\begin{eqnarray}
\label{an2}
u=e^{-c\xi/2}\left[A\cos(\alpha \xi)+B\sin(\alpha \xi)\right],
\end{eqnarray}
where $\alpha=[4F'(0)-c^2]/2$.
In any case the solution of (\ref{vv})  satisfies the boundary condition (\ref{cococo}) for any $A$ and $B$. So the boundary condition (\ref{cococo}) can be discarded. 

Because we have lost one boundary condition the solution of (\ref{86}) contains an arbitrary constant. And here our mechanical analogy helps us to understand that this (irrelevant) constant describes the translation of the solution in "time" $\xi$.

After we understood the true nature of the boundary conditions (\ref{bc}) for (\ref{86}),
the situation with the boundary conditions (\ref{bmp}) for Eq. (\ref{pp}) -
which are also exceptional - becomes obvious. 

In the vicinity of $u=1$,  Eq. (\ref{pp}) becomes asymptotically
\begin{eqnarray}
\label{pp7}
qq_{v}-cq=|F'(1)|v,
\end{eqnarray}
where again $v=1-u$, and $q\equiv dv/d\xi$.
Equation (\ref{pp7}) has two solutions 
\begin{subequations}
\label{twos}
\begin{alignat}{4}
q&=s_1v, \label{s1}\\
q&=s_2v, \label{s2}
\end{alignat}
\end{subequations}
where $s_{1,2}$ are given by (\ref{12}). Only the solution (\ref{s1}) satisfies the condition (\ref{coco}).
Hence
we realize that the boundary condition (\ref{coco2}) should be specified as follows: For $u\to 1$
\begin{eqnarray}
\label{coco22}
p=s_1(u-1).
\end{eqnarray}
(Not the difference between the proportionality sign in (\ref{cand}) and the equality sign in (\ref{coco22}).)
One can easily check up that our solution  (\ref{pmx}) satisfies this boundary condition. 

In the vicinity of $u=0$,  Eq. (\ref{pp}) becomes asymptotically
\begin{eqnarray}
\label{pp8}
pp_{u}+cp=-F'(0)u.
\end{eqnarray}
Equation (\ref{pp8}) has two solutions (we'll consider only the case $c^2> 4F'(0)$)
\begin{subequations}
\label{tw}
\begin{alignat}{4}
p&=s_3u,\label{333}\\
p&=s_4u,
\end{alignat}
\end{subequations}
where $s_{3,4}$ are given by (\ref{34}). Because both $s_3$ and $s_4$ are negative, both solutions will satisfy the boundary condition (\ref{cococo}). So we understood that the second boundary condition (\ref{coco3}) can be discarded -- it is satisfied automatically. 

Note that for the case considered explicitly in the previous Section
($F(u)$ being given by (\ref{fff}) and $c$ being given by (\ref{kkk}), obviously fulfilling the inequality $c^2>4a$ for any $n>0$ and $a\geq 0$) we get
\begin{subequations}
\begin{alignat}{4}
s_3&=-\frac{1}{\sqrt{n+1}},\\
s_4&=-a\sqrt{n+1}.
\end{alignat}
\end{subequations}
Looking at Eq. (\ref{pmx}) we realise that the solution (\ref{333}) is relevant in this case.

So finally, in the previous Section we were solving the first order ODE 
with the single boundary condition. The solution should exist for any given value of the speed $c$ and shouldn't contain an undefined constant.
We did finetune the speed $c$ in the previous Section for the solution of (\ref{pp2}) to be expressed in terms of elementary functions, and not for the solution to exist. 

Our analysis of (\ref{vv}) allows us to understand (partially)  the analytic properties of $p(u)$. If $c^2\geq 4F'(0)$, 
from (\ref{can2}) and (\ref{n2}) we can expect that $p(u)$ would be either a single - or a double-valued function. (Further analysis is necessary to decide which option is realised.) 

Looking at (\ref{an2}) (which corresponds to $c^2<4F'(0)$) on the other hand, we realize that $p(u)$ would be an 
infinitely-valued function. 
Note that if for physical reasons $u$ should be strictly nonnegative, such  values of speed are unacceptable, because from (\ref{an2}) follows that $u$ would acquire negative values in the vicinity of $\xi=+\infty$.
(Apart from recovering this well known property, we in this paper will say nothing about 
the front propagation speed selection, an issue which is both  important and complicated \cite{saarloos,ma}.)

\section{Conclusions}

We have obtained exact travelling wave solutions of the KPP equation for a wide class of the reaction terms and similar for the generalized Burgers–Huxley equation. Our method of integration of the second order differential equations is based on the choice of the original dependent variable as the new independent variable, and the derivative of the original dependent variable - as the new dependent variable. The boundary conditions for thus obtained first order differential equations were thoroughly studied.
All this gave us the opportunity to obtain the exact solutions mentioned above.

\begin{acknowledgments}
We are  grateful to  G. Gonzales, A. Gupta, M. Ma, B. Malomed, A. Mogilner, H. C. Rosu, C. Ou and R. Selvaraj for the insightful discussions.

\end{acknowledgments}
 
\begin{appendix}

\section{Exact solution of the generalized Duffing –Van der Pol equation}
\label{kudr}

Consider the generalized Duffing –Van
der Pol equation 
\begin{eqnarray}
\label{xxx}
X_{tt}+\left(a-bF_X\right)X_t=F(X),
\end{eqnarray}
where $X(t)$ is a deviation from the balance of the oscillator, $t$ is time,
$F(X)$ is the force field determined by the model, and  
$a,b$  are the parameters determining the generalized friction.
(Equation (\ref{xxx}) describes, in particular, travelling waves in a lossy transmission line with nonlinear capacitance \cite{kogan3}.)
Introducing new dependent variable 
\begin{eqnarray}
\label{px}
p=X_t
\end{eqnarray}
and considering $X$ as the independent variable,
we reduce (\ref{xxx}) to the first order ODE
\begin{eqnarray}
\label{yyy}
pp_X+\left(a-bF_X\right)p=F(X).
\end{eqnarray}
The particular solution of (\ref{yyy})
\begin{eqnarray}
\label{zzz}
p(X)=bF(X)
\end{eqnarray}
exists if
\begin{eqnarray}
\label{ab}
ab=1.
\end{eqnarray}
Looking back at Section \ref{two} we realize that the solution corresponds to the factorization
\begin{eqnarray}
F(X)=F(X)\cdot 1.
\end{eqnarray}

Substituting thus obtained function $p(X)$ into (\ref{px})  we obtain
the solution of (\ref{xxx})
\begin{eqnarray}
\label{i}
t=\int\frac{dX}{bF(X)}.
\end{eqnarray}

Consider $F(X)$ which has two adjacent zeros at 
$X_1$ and   $X_2$ ($F(X_1)=F(X_2)=0$), one of these points being the point of maximum of the potential $V(X)$ defined by the force field $F(X)$, the other - the point of minimum.  
Then (\ref{i})  describes explicitly the motion of the particle in the potential well $V(X)$ which starts in the infinite past at the potential maximum  and ends in the infinite future at the potential minimum. Finetuning of the friction (Eq. (\ref{ab})) is necessary for the absence of  oscillations in the vicinity of the potential minimum, which gives the opportunity to integrate (\ref{xxx}) in quadratures.

Let us be even more specific and  consider the Duffing –Van der Pol equation from Ref. \cite{19}
\begin{eqnarray}
\label{xx}
X_{tt}+\left(\alpha+\beta X^2\right)X_t=gX- X^3.
\end{eqnarray}
Equation (\ref{xx}) is the particular case of (\ref{xxx}) corresponding to
\begin{subequations}
\begin{alignat}{4}
F(X)&=gX- X^3,\\
a&=\alpha+\beta g/3 \\
b&=\beta/3.
\end{alignat}
\end{subequations}
Thus the condition (\ref{ab}) takes the form
\begin{eqnarray}
\alpha=3/\beta-\beta g/3; 
\end{eqnarray}
integration in (\ref{i}) can be easily performed and we reproduce the result of Ref. \cite{19}
\begin{eqnarray}
X(t)=\frac{\sqrt{g}}{\sqrt{1-e^{-2\beta gt/3}}}.
\end{eqnarray}

\section{Generalized Burgers–Huxley equation}
\label{bh}

The results obtained in Section \ref{two} for the generalized KPP equation (\ref{866})
can be easily generalised to the generalized Burgers–Huxley equation \cite{cherniha,kushner}
\begin{eqnarray}
\label{866c}
u_t-\lambda u^nu_x- u_{xx}=u\left(1-u^n\right)\left(u^n+a\right),
\end{eqnarray}
where $\lambda$ is a constant. 
The Burgers–Huxley equation
is known in various fields of applied mathematics. For example, it describes transport processes in systems when diffusion
and convection are equally important  and nonlinear reaction–diffusion processes too.
(An interesting further generalization of the model see in \cite{gupta}.)

For the travelling wave solution of (\ref{866c}) we obtain instead of (\ref{pp2}) the equation
\begin{eqnarray}
\label{ppc}
pp_{u}+\lambda u^np+cp=u\left(u^n-1\right)\left(u^n+a\right).
\end{eqnarray}
One can easily see that   the solution (\ref{pmx})
(and hence Eq. (\ref{mm})) is valid in the present case also,  same as the formula (\ref{kkk}).
The only difference is that Eq. (\ref{k}) in this case takes the form
\begin{eqnarray}
\label{kkkc}
P=\frac{\sqrt{\lambda^2+4(n+1)}-\lambda}{2(n+1)}.
\end{eqnarray}

If we put $a=0$ in the formula (\ref{kkk}), Eqs. (\ref{kkkc}) and (\ref{mm}) supply the solution of the generalized Huxley equation \cite{clarkson}
\begin{eqnarray}
\label{8}
u_t-\lambda u^nu_x- u_{xx}=u^{n+1}\left(1-u^n\right).
\end{eqnarray}

\section{Perturbation theory I}
\label{per}

Let us return to Eq. (\ref{pp}).
In this Appendix we'll present the expansion of the solution of the equation (and  hence of the solution of Eq. (\ref{86})) with respect to the powers of $1/c$ (meaningful if  $c\gg 1$).
Such expansion has the form
\begin{eqnarray}
\label{ex}
p(u)=\sum_{k=1}\frac{1}{c^k}p^{(k)}(u),
\end{eqnarray}
where
\begin{subequations}
\begin{alignat}{4}
p^{(1)}&=-F(u),\label{qp1}\\
p^{(k)}&=-\sum_{m=1}^{k-1}p^{(m)}p^{(k-m)}_u,\hskip .5cm k=2,3,\dots.
\end{alignat}
\end{subequations}

Let us restrict ourselves by the next to leading order approximation with respect to the parameter $1/c$. Substituting thus truncated  $p(u)$ into (\ref{cot}) we obtain 
\begin{eqnarray}
\frac{du}{d\xi}=-\frac{1}{c}F(u)\left(1+\frac{1}{c}\frac{dF(u)}{du}\right)
\end{eqnarray}
or, equivalently (with the accepted precision),
\begin{eqnarray}
\label{p}
\frac{d\xi}{c}=-\frac{du}{F(u)}+\frac{1}{c}\frac{dF(u)}{F(u)}.
\end{eqnarray}
Integrating (\ref{p}) we get
\begin{eqnarray}
\label{p7}
\frac{\xi}{c}=-\int\frac{du}{F(u)}+\frac{1}{c}\ln |F(u)|.
\end{eqnarray}

To make our life  simpler, further on
 consider only the particular case of $F(u)$ corresponding to the Fisher equation:
\begin{eqnarray}
F(u)=u(u-1).
\end{eqnarray}
Performing integration in (\ref{p7}) we obtain
\begin{eqnarray}
\label{eq}
\frac{\xi}{c}=\ln\left[\frac{1-u}{u}\right]+\frac{1}{c}\ln \left[u(1-u)\right].
\end{eqnarray}
In the leading order with respect to $1/c$ the solution of (\ref{eq}) is
\begin{eqnarray}
u(\xi)=&\frac{1}
{\exp\left(\xi/c\right)+1},
\end{eqnarray}
hence with the accepted precision we can rewrite (\ref{eq}) as
\begin{eqnarray}
\label{eq2}
\frac{\xi}{c}-\frac{1}{c}\ln\frac{\exp\left(\xi/c\right)}
{\left[\exp\left(\xi/c\right)+1\right]^2}=\ln\left[\frac{1-u}{u}\right].
\end{eqnarray}
Exponentiating, solving with respect to $u$ and again expanding with respect to $1/c$ 
we obtain 
after a bit of algebra 
\begin{eqnarray}
u(\xi)=\frac{1}
{\exp\left(\xi/c\right)+1}\nonumber\\
+\frac{1}{c}\frac{\exp\left(\xi/c\right)}
{\left[\exp\left(\xi/c\right)+1\right]^2}\ln\left\{\frac{\exp\left(\xi/c\right)}
{\left[\exp\left(\xi/c\right)+1\right]^2}\right\}.
\end{eqnarray}

\section{Perturbation theory II}
\label{2}

In Section \ref{two} we have solved Eq. (\ref{pp2})
 for $c=c_0$.
Let us now study the solution of the equation for $c$ close to $c_0$, considering $\delta c\equiv c-c_0$ as the small parameter and using the linear with respect to $\delta c$ approximation. To make life simpler, we'll consider only the case $n=1$. Presenting the solution  (for $n=1$) as 
\begin{eqnarray}
\label{p6}
p(u)=p_0(u)+\frac{\delta c}{\sqrt{2}} p_1(u)
\end{eqnarray}
and substituting (\ref{p6}) into (\ref{pp2}), for $p_1$ in linear approximation  we obtain equation
\begin{eqnarray}
\label{pp6}
u(u-1){p_1}_u+\left(2u-1+c_0\right)p_1+u(u-1)=0.
\end{eqnarray}
Integrating (\ref{pp6}) we obtain
\begin{eqnarray}
\label{p66}
p_1=e^{-Q(u)}
\left(C-\int e^{Q(u)}du\right),
\end{eqnarray}
where
\begin{eqnarray}
\label{q}
Q(u)\equiv\int \frac{2u-1+c_0}{u(u-1)}du,
\end{eqnarray}
and $C$ is an arbitrary constant. The lower limits of the integrals in (\ref{p66}) and (\ref{q}) can be taken arbitrary. Calculating the integral in (\ref{q}) we get
\begin{eqnarray}
\label{p77}
p_1(u)=\frac{u^{c_0-1}}{\left(1-u\right)^{c_0+1}}
\left(C-\int \frac{\left(1-u\right)^{c_0+1}du}{u^{c_0-1}}\right). 
\end{eqnarray}
Taking into account the boundary condition $p_1(1)=0$ we obtain
\begin{eqnarray}
\label{p8}
p_1(u)=\frac{u^{c_0-1}}{\left(1-u\right)^{c_0+1}}
\int_u^1 \frac{\left(1-v\right)^{c_0+1}dv}{v^{c_0-1}}. 
\end{eqnarray}

A question of the validity of the linear approximation presented above is quite delicate,
because the coefficient before 
${p_1}_u$ in  Eq. (\ref{pp6})  goes to zero when $u\to 0,1$. A good sign is that $p_1$ 
given by (\ref{p8}) has the asymptotic behavior  $p_1\sim (u-1)$ in the vicinity of $u=1$, in accordance with (\ref{coco22}). However,  in the vicinity of $u=0$, $p_1$ 
has the asymptotic behavior 
$p_1\sim u$, as it should be  in accordance with (\ref{tw}), only for $c_0>2$.
(For  $c_0<2$, $p_1\sim u^{c_0-1}$.)
So we would like to believe that 
for  $c_0>2$  the linear approximation is valid everywhere (for $0\leq u\leq 1$), and for
$c_0<2$  the linear approximation is valid everywhere apart from the vicinity of $u=0$.

\end{appendix}

\end{document}